\documentstyle[11pt,epsf]{article}
\input{psfig}
\catcode`@=11
\def\seceqaa{\@addtoreset{equation}{section}
\def\theequation{A\arabic{equation}}}
\def\seceqbb{\@addtoreset{equation}{section}
\def\theequation{B\arabic{equation}}}
\def\seceqcc{\@addtoreset{equation}{section}
\def\theequation{C\arabic{equation}}}
\catcode`@=12
\begin{document}
\begin{center}
{\Large \bf 
The Peculiarity of a  Negative Coordinate Axis in Dyonic Solutions 
of Noncommutative ${\cal N}=4$ Super Yang-Mills}
\vskip 0.1in
{Aalok Misra\\ 
Harish Chandra Research Institute,\\
Jhusi, Allahabad - 211 019, India\\email: aalok@mri.ernet.in}
\vskip 0.5 true in
\end{center}

\begin{abstract}

We show that in the neighborhood of a negative coordinate axis,  the $U(1)$
sector of the static dyonic solutions to the noncommutative $U(4)$ ${\cal N}=4$
Super Yang-Mills (SYM) can be consistently decoupled from the $SU(4)$ 
to {\it all orders in the noncommutativity parameter}. 
We show the above decoupling in two ways. First, we show 
the noncommutative dyon being the same as the commutative dyon, is a 
consistent solution to noncommutative equations of motion in the abovementioned
region of noncommutative space.  Second, as an example of 
decoupling of a non-null $U(1)$  sector, 
we also obtain a family of solutions with nontrivial 
$U(1)$ components for all components of the gauge 
field in the same region of noncommutative space. 
In both cases, the $SU(4)$ and $U(1)$ components separately 
satisfy the equations of motion.
\end{abstract} 

\section{Introduction}

Nonperturbative solutions in Super Yang-Mills (SYM) theories have been an 
important area of work. More recently, static dyonic solutions 
in $SU(N), N\geq3,4$ ${\cal N}=4$ SYM corresponding to planar and 
non-planar (respectively) string junctions have been
obtained in \cite{SI1,SI2} (See also references therein). Also, solitonic
solutions in  noncommutative SYM have been considered by others before 
(See \cite{hata1,hata2} and references therein). Now, the gauge group
$SU(N)$, in general is not allowed in noncommutative gauge
theories \cite{ncggps} 
(See \cite{wess} though, where the gauge parameter in
noncommutative gauge tranformations is allowed to depend on the usual $SU(N)$
gauge paramater and gauge fields. The explicit form of the map to all orders
in the noncommutativity parameter [which is what we are interested in,
in this paper] is quite involved and has not been worked out.). 
One of the questions that we address in this work is whether it is possible
to construct solutions to the equations of motion of a noncommutative
supersymmetric gauge theory in which one can decouple the $U(1)$ components
from the $SU$ components of the fields at least in some region of the
noncommutative space, {\it to all orders in the noncommutativity paramater}. 
There are the following two additional motivations 
for this work. The equality of the angular momentum and (null) 
quadrupole moment of the commutative and noncommutative monopoles of (S)U(2) 
$N=2$ SYM  up to $O(\theta^2)$ was shown in \cite{noncomcomeq} 
and in all of space. This makes it natural to pose the following question:
is it possible to construct solutions of a noncommutative supersymmetric
gauge theory that are characterized by quantities that receive no
noncommutativity corrections {\it to all orders in the noncommutativity
paramater}? Finally, it is of obvious interest to construct solutions
that are valid up to all orders in the noncommutativity paramater.

In this paper, we show that for noncommutative $U(4)$ ${\cal N}=4$ SYM, 
the static dyonic solutions in the neighborhood of a negative coordinate axis
of the noncommutative space, are such that one can decouple 
the $U(1)$ from the $SU(4)$, and construct
a 6-parameter family of noncommutative dyonic solutions (for a non-null 
$U(1)$ decoupling). Even though, the decoupling has been shown for a very
limited region of the noncommutative space, the reason the solutions are of
interest is because they have the same electric and magnetic charges as  and 
energy differing infinitesimally from
the commutative dyonic solutions of \cite{SI1}, {\it to all orders in the
noncommutativity parameter}. This points to the peculiar nature of the
the negative $x^3$ axis when constructing dyonic solutions to noncommutative
${\cal N}=4$ SYM, which requires further investigation.

The paper is organized as follows. In section 2, we discuss the basics
needed for getting static dyonic solutions in commutative ${\cal N}=4$ 
Super Yang-Mills (SYM). In section 3, 
we discuss the modifications in the techniques of section 2 to
get static dyonic solutions of noncommutative 
${\cal N}$=4 SYM. It has two subsections: (a short) 
3.1 on decoupling of a null $U(1)$ from the $SU(4)$, and 3.2 on
decoupling of a non-null $U(1)$ from the $SU(4)$. In section 4, we 
give a summary of and a discussion on the results obtained in the paper, 
and indicate possible future directions for work.

\section{Basics}

In this section, we give a brief review of the techniques given
in \cite{SI1} to obtain dyonic solutions in commutative $SU(4)$ ${\cal N}=4$
SYM.

It is advisable to write the equations of motion as 
if one were working in curved space-time.  The reason is that for 
solutions of the equations of motion, it is convenient to go to 
the stereographic coordinate system (See Fig 1: $z=e^{i\phi}tan{\theta\over2}$
) in which the metric is not
constant. The equations of motion for commutative $SU(4)$ ${\cal N}=4$ SYM
are given by:
\begin{eqnarray}
\label{eq:comeoms}
& &
{D_i}(\sqrt{g}g^{ii_1}g^{jj_1}{F}_{i_1j_1})=i\sqrt{g}\sum_J
[{D^j}{\Phi^J},{\Phi^J}];\nonumber\\
& &
{D_i}(\sqrt{g}g^{ij}{D}_j
{\Phi^J})=\sqrt{g}[{\Phi}^I,[{\Phi}^I,
{\Phi}^J]].
\end{eqnarray}
The component-form of the above equations is given in \cite{SI1}
which we won't be repeating here. Following the close parallel
with Skyrme  models\cite{projskr}, one uses the harmonic-map anzatz to
solve the equations of motion - the Higgs and gauge fields are 
assumed to have the following form:
\begin{eqnarray}
\label{eq:AzPhidefs}
& & {A_z}= {A_z}^a T^a 
=-(1+c_1-c_2-c_0)(r)(\partial P_0) P_0 + (c_2-c_1)(r)[P_1,\partial P_1] 
+ (1-c_2(r))[P_2,\partial P_2],\nonumber\\
& & {A_r}=0;\nonumber\\
& & {\Phi^I}={\Phi}^{I,a} T^a =
(b^I_2+b^I_0+b^I_1)(r)(P_0
-{1\over4}{\bf 1}_4) + (b^I_2+b^I_1)(r)(P_1-{1\over4}{\bf 1}_4)
+b^I_2(r)(P_2-{1\over4}{\bf 1}_4),\nonumber\\
& & 
\end{eqnarray}
$T^a$ being the generators of $SU(4)$, $I$ indexing the number of scalars
that are turned on, and for this paper taking values 1,2 and 3,
$c_{0,1,2}(r)$ and $b^I_{0,1,2}(r)$ being the `profile functions' of the 
gauge field and Higgs respectively, and $P_i$ the orthogonal set of 
${\bf CP}^3$ sigma model projectors defined as:
\begin{eqnarray}
\label{eq:defsprojs}
& & P_0={ff^\dagger\over f^\dagger f};\nonumber\\
& & P_j={(\bigtriangleup^j f)(\bigtriangleup^j f)^\dagger\over
|\bigtriangleup^j f|^2},\ {j=1,2},
\end{eqnarray}
where 
\begin{eqnarray}
\label{eq:Nablafdef}
& & \bigtriangleup f\equiv \partial f - f{f^\dagger\partial f
\over|f|^2};
\nonumber\\
& & f\equiv(1,\sqrt{3} z, \sqrt{3} z^2, z^3)^t
,  
\end{eqnarray}
and $\bigtriangleup^2f=\bigtriangleup(\bigtriangleup f)$.
They can be thought of as hermitian maps from $S^2$ to ${\bf CP}^3$. 
The explicit forms of the projectors $P_i$s can be worked out
using Mathematica or Maple.

The substitution of the harmonic map anzatz into the equations of motion
(written in components) either results in an equation of motion being
identically satisfied, or a system of 
non-linear second-order coupled differential equations
involving the Higgs and gauge field profile functions
that can be solved using techniques referred to in \cite{SI1}.
The solutions corresponding to (non-)planar string networks  
(for non-planar, assuming an infinitesimal non-planarity corresponding to
an infinitesimal deformation from the BPS nature of the planar solution)
were then obtained in \cite{SI1}.

\section{Noncommutative dyonic solutions}

In this section we derive a class of static noncommutative dyonic solutions
assuming  $r-z$ and $r-{\bar z}$ noncommutativity only. Thus, the
star product of two functions $f(r,z,{\bar z})$ and $h(r,z,{\bar z})$ will
be:
\begin{equation}
\label{eq:rzrzbarncdieuf}
f(r,z,{\bar z})*h(r,z,{\bar z})
=e^{{i\over2}[\Theta^{rz}(\partial_r^1\partial^2 - \partial_r^2\partial^1)
+\Theta^{r{\bar z}}(\partial_r^1{\bar\partial}^2 
- \partial_r^2{\bar\partial}^1)]}f(r_1,z_1,{\bar z}_1)
h(r_2,z_2,{\bar z}_2)|_{1=2}.
\end{equation}
We show that using stereogrpahic coordinates, in the neighborhood
of the South Pole, or equivalently the neighborhood of the negative $x^3$ axis,
one can consistently decouple in the equations of motion,
the $U(1)$ from the $SU(4)$ components of the Higgs and the
gauge fields.

The noncommutative equations of motion are given by:
\begin{eqnarray}
\label{eq:noncomeoms}
& &
\hat{D_i}*(\sqrt{g}*g^{ii_1}*g^{jj_1}*\hat{F}_{i_1j_1})=i\sqrt{g}*\sum_J
[\hat{D^j}*\hat{\Phi^J},\hat{\Phi^J}]_*;\nonumber\\
& &
\hat{D_i}*(\sqrt{g}*g^{ij}*\hat{D}_j*
\hat{\Phi^J})=\sqrt{g}*[\hat{\Phi}^I,[\hat{\Phi}^I,
\hat{\Phi}^J]_*]_*.
\end{eqnarray}
As we are considering only $r-z, r-{\bar z}$-noncommutativity,
the projectors of \cite{SI1} are still valid here. 
We assume the following anzatz for the gauge field $\hat{A_z}$, 
$\hat{A_r}$ and the Higgs field $\hat{\Phi^I}$:
\begin{eqnarray}
\label{eq:AzPhihatdefs}
& & \hat{A_z}= \hat{A_z}^a T^a +i\epsilon F_z(\Theta)
g_z(r)\hat{{\cal A}^0_z}{\bf 1}_4;\nonumber\\
& & \hat{A_r}=i\epsilon 
F_r(\Theta)g_r(r)\hat{{\cal A}^0_r}{\bf 1}_4;\nonumber\\
& & \hat{\Phi^I}=\hat{\Phi}^{I,a} T^a+i\epsilon F_\Phi(\Theta)g_\Phi^I(r)
\hat{{\cal\phi}^0}{\bf 1}_4,\nonumber\\
\end{eqnarray}
where $F_{z,r,\Phi}(\Theta)$ are functions of the noncommutativity
parameter $\Theta$ which vanish as $\Theta$ is set to zero and are bounded
as $\Theta\rightarrow\infty$, and
\begin{eqnarray} 
\label{eq:anztzSU}
& & \hat{A_z}^a
T^a=-(1+c_1-c_2-c_0)(r)*(\partial P_0) P_0 + (c_2-c_1)(r)*[P_1,\partial P_1] 
+ (1-c_2(r))*[P_2,\partial P_2],\nonumber\\
& & \hat{\Phi}^{I,a}T^a=
(b^I_2+b^I_0+b^I_1)(r)*(P_0
-{1\over4}{\bf 1}_4) + (b^I_2+b^I_1)(r)*(P_1-{1\over4}{\bf 1}_4)
+b^I_2(r)*(P_2-{1\over4}{\bf 1}_4)
,\nonumber\\
& & 
\end{eqnarray}
$c_{0,1,2}(r)$ and $b^I_{0,1,2}(r)$ being the `profile functions' of the 
gauge field and Higgs respectively as in (\ref{eq:AzPhidefs}).  
We have thus taken an anzatz in which the $U(4)$ 
fields have infinitesimal $U(1)$ components. We will
work up to $O(\epsilon)$. Hence, the noncommutative $U(1)$ field
strengths are the same as the commutative $U(1)$ field strengths.
There is no star product in the anzatz (\ref{eq:AzPhihatdefs}) in the $U(1)$ 
components of the fields because as argued
below, the star products can be dropped in the neighborhood of the South 
Pole in the $SU(4)$-components of the fields, and we assume
similar behavior for the $U(1)$ and $SU(4)$ components. The anzatz
for the $U(1)$ components that we choose in (\ref{eq:U1anztz}) below,
is such that if one were to put in 
star product between the radial and angular functions, then
it (the star product) can not be dropped in the neighborhood of the
South Pole.

As {\it expanding the star product in (\ref{eq:anztzSU}) will involve taking
derivatives of ($\partial) P_i$}, using the explicit 
forms of the projectors that can be worked out using Mathematica or
Maple,  one sees that 
{\it one can drop $(\partial)\partial P_i$ relative to $(\partial) P_i$ 
as one approaches the neighborhood of the South Pole}, as the former
is suppressed relative to the latter by terms of $O({1\over|z|^m}), m>0$ 
that vanish as $|z|\rightarrow\infty$ (i.e. in the 
neighborhood of the south pole).  {\it One thus has for the $SU$
part of $\hat{A_z},\hat{\Phi}$ the same expression as in
\cite{SI1}.} 
One can argue similarly that the * products in (\ref{eq:noncomeoms}) can also
be dropped (in the neighborhood of the South Pole).
We now discuss the decoupling from $SU(4)$, 
of a null $U(1)$ in {\bf 3.1} and non-null $U(1)$ in {\bf 3.2}.

\subsection{Decoupling of null $U(1)$}

Using (\ref{eq:AzPhihatdefs}), for null $U(1)$ components,
one can require consistently that the $SU(4)$ and $U(1)$ 
components, separately, satisfy the equations of motion, 
and one can show there is no mixing between the $SU(4)$ and $U(1)$ sectors.  
Hence, one can consistently set the $U(1)$ components to zero.
The entire analysis of \cite{SI1} to show the connection between (non-)planar 
stable (non-)BPS string junctions and dyon solutions, then goes through 
even in the neighborhood of the South Pole of the abovementioned 
noncommutative space if one notices and uses the following 
transformation that takes one from the North Pole to 
the South Pole (and is also a symmetry of the  equations of motion of the 
$SU(4)$ part of the solutions, as given in reference two of \cite{SI1}):
\begin{equation}
\label{eq:sym}
c_0\leftrightarrow c_2;\ b^I_0\leftrightarrow b^I_2.
\end{equation}
One also uses the fact that if $\hat{\Phi}$ is a solution then
$-\hat{\Phi}$ is also a solution to the equations of motion.

\subsection{Decoupling of non-null $U(1)$}

Here we discuss in detail, the decoupling of non-null $U(1)$ components 
of the Higgs and the gauge fields from their $SU(4)$ components in the 
neighborhood of the South Pole of the noncommutative space.
To show the decoupling of a non-null $U(1)$ from $SU(4)$,
we choose the following anzatz for the $U(1)$ components (of the 
Higgs and gauge fields):
\begin{eqnarray}
\label{eq:U1anztz}
& & \hat{{\cal\phi}^0},\hat{{\cal A}_{z}^0}=
Det[(P_0-a_{\Phi,z} {\bf 1}_4)(P_1-M_{\Phi,z})(P_2-M_{\Phi,z})]+
{\rm cc}
\nonumber\\
& & \hat{A_r^0}= z Det[(P_0-a_{r} {\bf 1}_4)(P_1-M_{r})(P_2-M_{r})]+{\rm
    cc}.\nonumber\\
\end{eqnarray}
where
\begin{equation}
\label{eq:M_idef}
M_i=\left(\begin{array}{cccc}\\
0&b_i&0&0\\
c_i&0&0&0\\
0&0&0&e_i\\
0&0&f_i&0\\
\end{array}\right),
\end{equation}
where $i\equiv r,z,\Phi$, and the parameters $a_i, b_i, c_i,  e_i$ and $f_i$
have been introduced to get non-zero determinants using the projectors
and some constant matrices; the parameters  get constrained later.
We now assume that one approaches the South pole along any path very close to
$\phi(\equiv$ azimuthal angle)$={\pi\over2}$. This implies that $Re z$ is
finite, but $Im z$ approaches infinity. Hence, one sees that 
$\partial{\bar\partial}\hat{{\cal A}_r^0}$ (as shown in  
(\ref{eq:laphatArzinf}))  and $(\partial-{\bar\partial}) 
(\hat{\phi}^0,\hat{{\cal A}_z^0})$ 
(as shown in  (\ref{eq:diffzinf})) are real. 
If 
\begin{equation}
\label{eq:cond1}
b_i=f_i,\ c_i=e_i,
\end{equation}
then  $\partial\biggl((1+|z|^2)^2(\partial+{\bar\partial})(\hat{\phi}^0,
\hat{{\cal A}_z^0})\biggr)$ 
(as shown in (\ref{eq:derz^4}))
will be finite and real as $Im\ z\rightarrow\infty$. These considerations
become useful when one solves the equations of motion for the $U(1)$ components
as done in (\ref{eq:U1eqns1}) and (\ref{eq:U1eqns2}) in the neighborhood of
the South Pole.

From power (of $|z|$ or $Im z$) counting 
arguments using expressions given in the  appendix, one 
can require consistently that the $SU(4)$ and $U(1)$ 
components, separately, satisfy the equations of
motion, and one can show there is no mixing between the $SU(4)$ and 
$U(1)$ sectors. As the decoupling for non-null $U(1)$ is non-trivial,
to illustrate the point, we consider the example of $\hat{F_{rz}}$.
We see that up to $O(\epsilon)$:
\begin{equation}
\label{eq:decplngod1ex}
\hat{F_{rz}}^{U(4)}=\hat{F_{rz}}^{SU(4)}
+\hat{F_{rz}}^{U(1)}-i[\hat{A_r}^{U(1)},\hat{A_z}^{SU(4)}]_*,
\end{equation}
Now,
\begin{eqnarray}
\label{eq:decplgod2ex}
& & [\hat{A_r}^{U(1)},\hat{A_z}^{SU(4)}]_*=i\Theta^{rz}\Biggl(
\partial_r\hat{A_r}^{U(1)}\partial\hat{A_z}^{SU(4)}-
\partial\hat{A_r}^{U(1)}\partial_r\hat{A_z}^{SU(4)}\Biggr)
\nonumber\\
& & 
+i\Theta^{r{\bar z}}\Biggl(
\partial_r\hat{A_r}^{U(1)}{\bar\partial}\hat{A_z}^{SU(4)}-
{\bar\partial}\hat{A_r}^{U(1)}
\partial_r\hat{A_z}^{SU(4)}\Biggr)
+O(\Theta^2).
\end{eqnarray}
Using expressions in the appendix, one sees that:
\begin{eqnarray}
\label{eq:decplgod3ex}
& & \partial_r\hat{A_r}^{U(1)}\equiv O(1),\ \partial\hat{A_r}^{U(1)},
{\bar\partial}\hat{A_r}^{U(1)}\equiv O(1);\nonumber\\
& & \partial\hat{A_z}^{SU(4)}\equiv O({1\over|z|^3})\times M,\
{\bar\partial}\hat{A_z}^{SU(4)}\equiv O({1\over|z|^4})\times
{\bf 1}_4,\
\partial_r\hat{A_z}^{SU(4)}\equiv O({1\over|z|^2})\times 
M;\nonumber\\
& & \partial_r\hat{A_z}^{U(1)}\equiv O(1),\
\partial\hat{A_z}^{U(1)},{\bar\partial}\hat{A_z}^{U(1)}\equiv
O({1\over|z|^2})({\rm relevant\ 
for\ }[\hat{A_r}^{U(1)},\hat{A_z}^{U(1)}]_*),\nonumber\\
& & 
\end{eqnarray}
where
\begin{equation}
\label{eq:defM}
M\equiv\left(\begin{array}{cccc}\\
0&1&0&0\\
0&0&1&0\\
0&0&0&1\\
0&0&0&0\\
\end{array}\right).
\end{equation}
Thus
\begin{equation}
\label{eq:decplgod4ex}
\hat{F_{rz}}^{U(1)}\equiv O(1);\hat{F_{rz}}^{SU(4)}
\equiv O({1\over|z|^2})\times M,
\end{equation}
and in Table 1, we show which terms can be dropped relative to
which terms as $Im z\rightarrow\infty$.
\begin{table} [htbp]
\centering
\caption{Consequences of power counting done in (\ref{eq:decplgod3ex})}
\begin{tabular}{|c|c|} \hline
Term that can be dropped & relative to \\ \hline
$\partial_r\hat{A_r}^{U(1)}\partial
\hat{A_z}^{SU(4)}$ & $\hat{F_{rz}}^{SU(4)}$ \\
$\partial\hat{A_r}^{U(1)}\partial_r\hat{A_z}^{SU(4)}$ &
$\hat{F_{rz}}^{SU(4)}$\\
$\partial_r\hat{A_r}^{U(1)}{\bar\partial}\hat{A_z}^{SU(4)}$
& $\hat{F_{rz}}^{U(1)}$\\
${\bar\partial}\hat{A_r}^{U(1)}\partial_r\hat{A_z}^{SU(4)}$&
$\hat{F_{rz}}^{SU(4)}$\\ \hline
\end{tabular}
\end{table}
Similar reasoning can be used to argue that terms of higher order
in $\Theta$, can also be dropped relative to the commutative
counterparts. Hence, as $Im z\rightarrow\infty$,
\begin{equation}
\label{eq:decplgod5}
\hat{F_{rz}}^{U(4)}\sim \hat{F_{rz}}^{SU(4)}+\hat{F_{rz}}^{U(1)},
\end{equation}
showing the stated decoupling.  

By requiring that the $SU$ components of the Higgs 
and gauge fields still satisfy the commutative
equations of motion (having dropped the star product at the 
South pole), the $U(1)$ components should satisfy these equations:
\begin{eqnarray}
\label{eq:U1eqns1}
& & {dg_z\over dr}(\partial-{\bar\partial})\hat{{\cal A}^0_z}-
2g_r\partial{\bar\partial}\hat{{\cal A}^0_r}
=0;\nonumber\\
& & {d^2g_z\over dr^2}
\hat{{\cal A}^0_z} - g_r\partial\hat{{\cal A}^0_r} 
+{g_z\over2r^2}
\partial\biggl((1+|z|^2)^2(\partial+{\bar\partial})\hat{A_z}
\biggr)=0
\nonumber\\
& & {1\over r^2}{d\over dr}(r^2{dg_\Phi^I\over dr})\hat{{\cal\phi}^0}
+{(1+|z|^2)^2\over r^2}g_\Phi^I
\partial{\bar\partial}\hat{{\cal\phi}^0}=0.
\end{eqnarray}
The * product has been dropped as it will generate $O({1\over{|z|^m}})$-
corrections (relative to $O(1)$),
which can be dropped in  the neighborhood of the 
south pole. We now discuss  a family of dyonic
solutions to (\ref{eq:U1eqns1}).

One gets (setting $c_r=e_r$ and $b_r=f_r$ for convenience):
\begin{eqnarray}
\label{eq:U1eqns2}
& & {dg_z\over dr}
+2g_r{[(b_rc_r)^3  
(2c_r - b_r)]\over(b_zc_z)^3(c_z-b_z)}
=0
\nonumber\\
& &  2{d^2g_z\over dr^2} (-1+a_{z})a^3_{z}
(b_{z}c_{z})^4 -
g_r(-1+a_r)a_r^3(b_rc_r)^4
+{
g_z\over r^2}(3(-1+a_z)a_z^3b_z^2c_z^2(b_z-c_z)^2)
=0\nonumber\\
& &  2{1\over r^2}{d\over dr}(r^2{dg_\Phi^I\over dr})(-1+a_{\Phi})a^3_{\Phi}
(b_{\Phi}c_{\Phi})^4
\nonumber\\
& & 
-{6\over r^2}|z|^2g_\Phi^I(-1+a_{\Phi,z})a_{\Phi}^3
(c_{\Phi}^4b_{\Phi}^2+b_{\Phi}^4c_{\Phi}^2)=0.
\end{eqnarray}
By eliminating $\hat{{\cal A}^0_r}$ from the first two equations, one gets the
following differential equation for $g_z$:
\begin{eqnarray}
\label{eq:diffeqgz}
& & 
{d^2g_z\over dr^2} 
+{dg_r\over dr}{(-1+a_r)a_r^3(b_rc_r)^4\sqrt{3}
(-1+a_z)a_z^3[(b_zc_z)^3(c_z-b_z)]\over
4(-1+a_z)^2a^6_z(b_zc_z)^4[b_r^3 c_r^3 
\sqrt{3} (2c_r - b_r)]}
\nonumber\\
& & 
+
{g_z\over r^2}(3(-1+a_z)a_z^3{(b_z-c_z)^2\over2 b_z^2c_z^2})=0
\end{eqnarray}
The equation (\ref{eq:diffeqgz}) is of the form
\begin{equation}
\label{eq:bessleqn}
r^2 y''(r) + \alpha r^2 y'(r) + \beta  y(r)=0,
\end{equation}
the solution to which (using maple) is given by:
\begin{eqnarray}
\label{eq:besselsolnnotZ}
& & 
y(r)=
{\it C_1}\sqrt{r}
{\it I}_{{1\over2}\sqrt{1-4\beta }}({1\over2}\alpha r)
{e^{-{1\over2}\,\alpha r}}\nonumber\\
& & +
{\it C_2}
\sqrt {r}{\it I}_{-{1\over2}\sqrt{1-4\beta }}
({1\over2}\alpha r){e^{-{{1\over2}\,\alpha r}}}
\end{eqnarray}
for ${1\over2}\sqrt{1-4\beta }\not\in{\bf Z}$and
\begin{eqnarray}
\label{eq:besselsolnZ}
& & y(r)=
{\it C_1^\prime}\sqrt{r}
{\it I}_{{1\over2}\sqrt{1-4\beta }}({1\over2}\alpha r)
{e^{-{1\over2}\,\alpha r}}\nonumber\\
& & +
{\it C_2^\prime}
\sqrt {r}{\it K}_{{1\over2}\sqrt{1-4\beta }}
({1\over2}\alpha r){e^{-{{1\over2}\,\alpha r}}}
\end{eqnarray}
for ${1\over2}\sqrt{1-4\beta }\in{\bf Z}$. Lets consider the two cases 
separately. 

For the former, to make sure that the solution goes to zero as
$r\rightarrow
\infty$, one sets:
\begin{equation}
\label{eq:condCs}
C_1=-C_2.
\end{equation}
Even though this is a singular solution (at $r=0$), as we shall see,
it is a finite energy solution.

The derivative of the solution involving $I_{\pm\nu}(\alpha r)$ is given by:
\begin{eqnarray}
\label{eq:derBesselI}
& & {d\over dr}
\Biggl(e^{-{\alpha r\over2}}\sqrt{r}\biggl(
I_{{1\over2}\sqrt{1-4\beta}}({\alpha r\over2})-I_{-{1\over2}
\sqrt{1-4\beta}}({\alpha r\over2})\biggr)
\Biggr)\nonumber\\
& & 
={e^{{-\alpha r\over2}}\over4\sqrt{r}}\biggl(\alpha 
r I_{-1+{1\over2}\sqrt{1-4\beta}}(
{\alpha r\over2}) +\alpha r I_{1+{1\over2}\sqrt{1-4\beta}}({\alpha r\over2})
+2(-1+\alpha r)I_{-{1\over2}\sqrt{1-4\beta}}({\alpha r\over2})\nonumber\\
& & -\alpha r I_{-1-{1\over2}\sqrt{1-4\beta}}({\alpha r\over2})
-\alpha r I_{1-{1\over2}\sqrt{1-4\beta}}({\alpha r\over2})
-2(-1+\alpha r)I_{{1\over2}\sqrt{1-4\beta}}({\alpha r\over2})\biggr).
\end{eqnarray}
For the purpose of getting finite energy, as we will very shortly
see, one requires to impose that $r^2({dy(r)\over dr})^2\rightarrow0$
as  $r\rightarrow\infty$, 
and that it is finite at $r=0$.
Using the series expansions around $r=0$ and the asymptotic expansions
for $I_\nu(r)$, the former is satisfied identically, 
and the latter is satisfied if one imposes:
\begin{equation}
\label{eq:constbeta}
\sqrt{1-4\beta}<1 \Leftrightarrow\ 0<\beta<{1\over4}.
\end{equation}

The  solution for $g_\Phi^I(r)$ that one gets is: 
\begin{equation}
\label{eq:gPhisoln}
g_\Phi^I(r)=0.
\end{equation}
Hence, the final result for the $U(1)$ components of the Higgs and
gauge fields, in the neighborhood of the South Pole (approaching it
along $\phi$ close to ${\pi\over2}$) are:
\begin{eqnarray}
\label{eq:final1}
& & \hat{A}^0_z(r,z)=i \epsilon
f_z(\Theta)\sqrt{r}{e^{-{1\over2}\,\alpha r}}\biggl(
{\it I}_{{1\over2}\sqrt{1-4\beta }}({1\over2}\alpha r)-
{\it I}_{-{1\over2}\sqrt{1-4\beta }}({1\over2}\alpha r)\biggr)
\nonumber\\
& & \times
Det[(P_0-a_z{\bf 1}_4)(P_1-M_r)(P_2-M_r)] +
{\rm cc});\nonumber\\
& & \hat{A^0_r}=i \epsilon f_r(\Theta)\nonumber\\
& & \times{e^{{-\alpha r\over2}}\over4\sqrt{r}}\biggl(\alpha r
I_{-1+{1\over2}\sqrt{1-4\beta}}(
{\alpha r\over2})
+\alpha r I_{1+{1\over2}\sqrt{1-4\beta}}({\alpha r\over2})
+2(-1+\alpha r)I_{-{1\over2}\sqrt{1-4\beta}}({\alpha r\over2})\nonumber\\
& & -\alpha r I_{-1-{1\over2}\sqrt{1-4\beta}}({\alpha r\over2})
-\alpha r I_{1-{1\over2}\sqrt{1-4\beta}}({\alpha r\over2})
-2(-1+\alpha r)I_{{1\over2}\sqrt{1-4\beta}}({\alpha r\over2})\biggr)
\nonumber\\
& & \times(z 
Det[(P_0-a_z{\bf 1}_4)(P_1-M_r)(P_2-M_r)] +
{\rm cc});\nonumber\\
& &
\hat{\Phi^0}=0,
\end{eqnarray}
where 
\begin{eqnarray}
\label{eq:defs}
& & \beta\equiv(3(-1+a_z)a_z^3{(b_z-c_z)^2\over2
  b_z^2c_z^2})<0;\nonumber\\
& & \alpha\equiv{(-1+a_r)a_r^3(b_rc_r)^4\sqrt{3}
(-1+a_z)a_z^3[(b_ze_z)^3(c_z-b_z)]\over
4(-1+a_z)^2a^6_z(b_zc_z)^4[b_r^3 c_r^3 
\sqrt{3} (2c_r - b_r)]};\nonumber\\
& & 0<a_z<1;\ c_{z,r}=e_{z,r};\
b_{z,r}=f_{z,r}.
\end{eqnarray}

Lets consider the energy of the above solution. It is given by (we
follow the conventions of \cite{SI1}):
\begin{eqnarray}
\label{eq:energy1}
& & E=-{1\over4\pi}\int r^2 dr {dz d{\bar z}\over(1+|z|^2)^2}\
 Tr\biggl[(\hat{D_i}*\hat{\Phi})^2
+{1\over2}(\hat{F_{ij}})^2\ _*\biggr]\nonumber\\
& & =E_{SU(4)}+E_{U(1)}.
\end{eqnarray}
But, in the neighborhood of the South Pole, where one can drop the * product,
\begin{eqnarray}
\label{eq:energy2}
& & E_{U(1)}
=-\int r^2 dr {dz d{\bar
    z}\over{(1+|z|^2)^2}}\biggl[(\partial_r
\hat{{\cal A}^0}_z -\partial\hat{{\cal A}_r^0})^2+
(\partial_r
\hat{{\cal A}^0}_z +{\bar\partial}\hat{{\cal A}_r^0})^2\nonumber\\
& & +((\partial+{\bar\partial})\hat{{\cal A}_z^0})^2\biggr]\geq0.
\end{eqnarray}
Hence, in the neighborhood of the South Pole, 
even though one is dealing with non-BPS solutions, one gets the
following inequality:
\begin{equation}
\label{eq:energy3}
E_{U(4)}\geq E_{SU(4)},
\end{equation}
with saturation of the inequality occuring for null $U(1)$ component.
Substituting (\ref{eq:final1}) in the $Im\ z\rightarrow\infty$ limit into
(\ref{eq:energy2}), and integrating in the neighborhood of the South
Pole as described earlier, i.e., around $\phi={\pi\over2}$ and
$\theta\in[\pi-\epsilon,\pi], \lim_{\epsilon\rightarrow0}$, 
one gets an infinitesimal result, provided one gets  a finite result 
from the $r$ integration. The $r$-integral should 
converge to get a finite-energy solution.  One sees that one has to evaluate:
\begin{eqnarray}
\label{eq:rintfinite}
& & E_{U(1)}\sim
\int_0^\infty dr r^2 {e^{-\alpha r}\over16r}\biggl(\alpha r
I_{-1+{1\over2}\sqrt{1-4\beta}}(
{\alpha r\over2})
+\alpha r I_{1+{1\over2}\sqrt{1-4\beta}}({\alpha r\over2})
+2(-1+\alpha r)I_{-{1\over2}\sqrt{1-4\beta}}({\alpha r\over2})\nonumber\\
& & -\alpha r I_{-1-{1\over2}\sqrt{1-4\beta}}({\alpha r\over2})
-\alpha r I_{1-{1\over2}\sqrt{1-4\beta}}({\alpha r\over2})
-2(-1+\alpha r)I_{{1\over2}\sqrt{1-4\beta}}({\alpha
  r\over2})\biggr)^2.\nonumber\\
\end{eqnarray}
We have not been able to evaluate the above integral. But the integrand
for $0<\beta<{1\over4}$ is analytic for all values of
$r\in[0,\infty)$, and it vanishes at $r=0$ and $r\rightarrow\infty$. We
hence assume that it is convergent corresponding to a finite energy
solution. We have verified the same numerically using mathematica for
several values of $\beta<{1\over4}$.

Let us now consider the case ${1\over2}\sqrt{1-4\beta}\in{\bf Z}$. In
that case in order to get a solution that vanishes as
$r\rightarrow\infty$, one sets $C_1^\prime=0$ in
(\ref{eq:besselsolnZ}).

One then requires:
\begin{eqnarray}
\label{eq:finenergZ}
& & E_{U(1)}\sim \int_0^\infty 
dr r^2
{e^{-\alpha r}\over16r}
\biggl(-\alpha r 
K_{-1+{1\over2}\sqrt{1-4\beta}}({\alpha r\over2}) 
- 2 (-1 + \alpha r) K_{{1\over2}\sqrt{1-4\beta}}({\alpha r\over2})  
\nonumber\\
& & - 
   \alpha r K_{1 + {1\over2}\sqrt{1-4\beta}}({\alpha r\over2})\biggr)^2<\infty.
\nonumber\\
& & 
\end{eqnarray}
By using the series expansion of $K_\nu(x)$ about $x=0$ (See
\cite{Bateman}), one sees that one can not find 
a $\beta: {1\over2}\sqrt{1-4\beta}\in{\bf Z}$ that will satisfy 
(\ref{eq:finenergZ}) because of the singular nature of the integrand
at $r=0$.

Now, lets turn to calculating the commutative Higgs 
and gauge field: $\Phi^I,A_{r,z,{\bar z}}$. As
we have thus far continued working to all orders in the
noncommutativity parameter $\Theta$, one hence requires an all-order (in
$\Theta$) Seiberg-Witten map\cite{SW,Liu}. There is a conjecture for that as
given in \cite{Liu}. We will not explicitly construct the SW map, but
conjecture that in the neighborhood of the negative $x^3$ axis (or the
neighborhood of the South Pole of the Riemann Sphere, approaching it
the way described in the paper), the noncommutative hatted fields are the
same as the commutative unhatted fields.  Noting that as 
$Im z\rightarrow\infty$, $\hat{A_{z,{\bar z}}}^{SU(4)},
\partial,{\bar\partial}\hat{\Phi}^{SU(4)},
\partial\hat{A_z}^{U(1)}$, $\partial^2\hat{A_r}^{U(1)}\rightarrow0$,
and dropping terms of $O(\epsilon^2)$ in $\hat{A_{r,z}}^{U(1)}$ in
equations (3.3), (3.4)  and (3.7) for the Higgs and equations  (A3), (A6) and
(A8) for the gauge field of  \cite{hatath2}, the conjecture can be  
explicitly verified to $O(\Theta^2)$.  Hence, in the neighborhood of 
the South Pole, approaching it along an azimuthal angle close to ${\pi\over2}$
\begin{eqnarray}
\label{eq:noncomcomequiv}
& & \hat{\Phi}=\hat{\Phi}^a T^a=\Phi^a T^a;\nonumber\\
& & \hat{A_i}=\hat{A_i^a} T^a + \hat{A_i^0}{\bf 1}_4=A_i^a T^a +
A_i^0{\bf 1}_4.
\end{eqnarray}
The electric charge is then the same as that given in \cite{SI1}. The
magnetic charge will receive a $U(1)$ contribution proportional to 
$(\partial+{\bar\partial}){A_z}(z,{\bar z},r\rightarrow\infty)$, which using
(\ref{eq:final1}), vanishes. Hence, the noncommutative dyon in the
neighborhood of the South Pole, carries the same electric and magnetic
charges as the commutative dyon of \cite{SI1}. Their energy will be
more than that of the commutative dyons in the neighborhood of the 
South Pole, though by an infinitesimal amount. As the equality of the
commutative and noncommutative electric and magnetic charges is valid
only in the neighborhood of the South Pole, these noncommutative dyonic
solutions are likely to be stable.

\section{Summary and Discussion}

To summarize, we have shown that noncommutative (with $r-z$ and
$r-{\bar z}$ noncommutativity) $U(4)$ SYM possesses static dyonic 
solutions in the neighborhood of the South Pole of the 
Riemann sphere or equivalently the negative $x^3$ axis, 
that either are the same as the static commutative $SU(4)$ dyonic
solutions of \cite{SI1} (corresponding to {\it decoupling of a null $U(1)$
from the $SU(4)$}) or, have the same electric and magnetic charges as 
the commutative dyons and differ infinitesimally in their energy (corresponding 
to {\it decoupling of a non-null $U(1)$ from the $SU(4)$}), 
to {\it all orders in the noncommutativity parameter}. 
We have thus shown the decoupling of the $U(1)$ sector, both for null and 
infinitesimal $U(1)$ components in the neighborhood of the negative $x^3$ axis. 
This decoupling to $O(\Theta)$ for the whole of space, was 
also seen in the context of noncommutative monopole solutions in 
SYM in \cite{hata1}.  The noncommutative hatted fields are
the same as the commutative unhatted fields (we conjecture to all orders in the
noncommutativity parameter and show explicitly up to $O(\Theta^2)$).  
Thus, we see that the neighborhood of the 
negative $x^3$ axis, acts as ``denoncommutatifier". 
It will be interesting to study the same problem using noncommutative ADHMN
construction (See \cite{NCADHMN} and references therein), and 
to be able to establish the existence of such
solutions using techniques of \cite{Taubes} using Morse theory. 
Also, finding similar solutions with $z-{\bar z}$ noncommutativity will be nice.

\section*{Acknowledgements} 

We thank A.Kumar for useful discussions. We
also thank K.L.Panigrahi, R.R.Nayak, S.S.Pal and G.Bhattacharya 
for bringing to our notice references \cite{SI1}, \cite{Liu} and \cite{ncggps}.

\appendix
\section*{Appendix A}
\setcounter{equation}{0}
\seceqaa

In this appendix, we summarize the asymptotic (w.r.t to $z$)
expressions for  $A^{U(1)}_{r,z}$
and some  expressions involving them necessary to obtain the
$U(1)$ components of the gauge field and Higgs. We have used
mathematica for arriving at the expressions given in this appendix.

Using (\ref{eq:U1anztz}), one can explicity write out the expressions for
$\hat{{\cal\phi}^0},\hat{A_{z}^0}=
Det[(P_0-a_{\Phi,z}
{\bf 1}_4)(P_1-M_{\Phi,z}).(P_2-M_{\Phi,z})] +{\rm cc}$ and 
$\hat{A_r^0}=z Det[(P_0-a_{r}
 {\bf 1}_4)(P_1-M_{r}).(P_2-M_{r})]+{\rm cc}$, using mathematica,
 but because of the length of the expressions, we will give only
 their asymptotic forms.

One gets as $Im z\rightarrow\infty$:
\begin{equation}
\label{eq:hatPhiAzzinf}
\hat{\cal\phi}^0,\hat{A_z^0}\sim 2(-1+a_{\Phi,z})a^3_{\Phi,z}
(b_{\Phi,z}c_{\Phi,z}e_{\Phi,z}f_{\Phi,z})^2.
\end{equation}

Simlarly, as $z\rightarrow\infty$, one gets:
\begin{equation}
\label{eq:hatArzinf}
\hat{A_r^0}\sim2(-1+a_r)a_r^3(b_rc_re_rf_r)^2Re z.
\end{equation}
Hence, we see that if $z\rightarrow\infty$ as $z=(Re z)_0+i\infty$,
where $(Re z)_0$ is finite, then $\hat{A_r}$ is non-singular, else it
is singular.

One sees that as $z\rightarrow\infty$,
\begin{equation}
\label{eq:dhatArzinf}
\partial\hat{{\cal A}^0_r}\sim(-1+a_r)a_r^3(b_re_rc_rf_r)^2.
\end{equation}

For the purpose of calculation of energy and  imposing it to be
finite,
one requires to calculate
${\bar\partial}\hat{\cal A}^0_r$, and its asymptotic expression is given by:
\begin{equation}
\label{eq:dervbarhatArzinf}
{\bar\partial}\hat{\cal A}_r^0\sim(-1+a_r)a^3_r b_r^2 c_r^2 e_r^2 f_r^2 
\end{equation}

One also needs to evaluate $\partial{\bar\partial}\hat{{\cal A}^0_r}$
as $z\rightarrow\infty$, and is given by:
\begin{equation}
\label{eq:laphatArzinf}
\partial{\bar\partial}\hat{{\cal A}^0_r}
\sim(-1+a_r)a_r^3
{b_r^2 c_r e_r^2 f_r 
(\sqrt{3} c_r - \sqrt{3} f_r){\bar z}^2+\sqrt{3}(b_re_rc_r)^2f_rz^2\over|z|^4}
\end{equation}
Also, one one sees that:
\begin{equation}
\label{eq:laphatPhiAzzinf}
\partial{\bar\partial}(\hat{\cal\phi}^0,\hat{A_z^0})
\sim-6(-1+a_{\Phi,z})a_{\Phi,z}^3
{(((c_{\Phi,z}e_{\Phi,z})^2b_{\Phi,z}f_{\Phi,z}+(b_{\Phi,z}f_{\Phi,z})^2c_{\Phi,z}
e_{\Phi,z})\over|z|^2}.
\end{equation}

For the purpose of solving for the $U(1)$ components of the gauge
field, and for getting the magnetic charge of the dyon, one needs to
evaluate $(\partial+{\bar\partial}){\cal A}^0_z$, and 
as $z\rightarrow\infty$, one gets:
\begin{eqnarray}
\label{eq:dersumzinf}
& & (\partial+{\bar\partial})\hat{{\cal A}^0_z}\sim
{b_z c_z^2 e_z f_z^2(\sqrt{3} b_z  - \sqrt{3} e_z){\bar z}^2
+(c_z f_z b_z^2(-e_z f_z\sqrt{3}e_z+ c_z e_z\sqrt{3} e_z)
-
3c_zf_zb_ze_zc_ze_z)z^2
\over|z|^4}
\nonumber\\
& & 
\end{eqnarray}

Two other quantities that one
needs to calculate are:
$(\partial-{\bar\partial})(\hat{\Phi},\hat{A_z})$
and
$ \partial\biggl((1+|z|^2)^2(\partial+{\bar\partial})
(\hat{\Phi},\hat{A_z})\biggr)$.
The former as $z\rightarrow\infty$, is given by:
\begin{equation}
\label{eq:diffzinf}
(\partial-{\bar\partial})(\hat{{\cal\phi}^0},\hat{{\cal A}_z^0})\sim
{\sqrt{3}(-1+a_i)a_i^3\biggl[(b_ie_i)^2c_if_i(c_i-f_i)
{\bar z}
+(-b_ic_i^2e_i^2f_i(f_i-b_i)+ e_ib_i^2c_if_i^2(c_i-e_i)
)z\biggr]
(z-{\bar z})\over|z|^4},
\end{equation}
The latter as $z\rightarrow\infty$ is given by:
\begin{eqnarray}
\label{eq:derz^4}
& & \partial\biggl((1+|z|^2)^2(\partial
+{\bar\partial})(\hat{{\cal\phi}^0},\hat{{\cal A}_z^0})\biggr)\sim
\nonumber\\
& & -2(-1+a_i)a_i^3\biggl(
\sqrt{3}[c_i^2e_i^2b_if_i(-f_i+b_i)+b_i^2f_i^2c_ie_i(c_i-e_i)]z
-(3b_if_i(c_ie_i)^2+3c_ie_i(b_if_i)^2)\nonumber\\
& & +{3b_ic_ie_if_i(b_ie_i+c_if_i){\bar z}^4\over|z|^4}\biggr)
\end{eqnarray}

\newpage
\begin{figure}[htbp]
\vskip -1in
\centerline{\mbox{\psfig{file=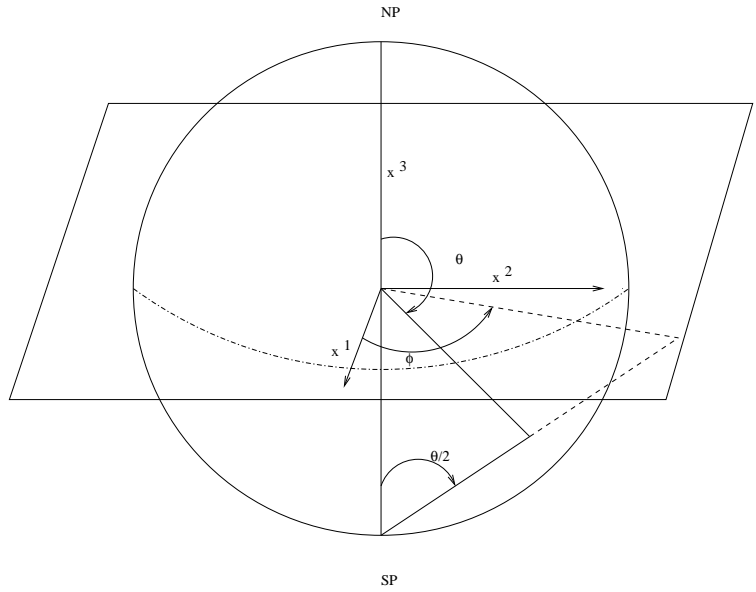,width=0.8\textwidth}}}
\caption{The stereographic coordinate system}
  \end{figure}
\end{document}